# Consecutive magnetic phase diagram of UCoGe-URhGe-UIrGe system

Jiří Pospíšil[a,b,*], Yoshinori Haga[a], Atsushi Miyake[c], Shinsaku Kambe[a], Naoyuki Tateiwa[a], Yo Tokunaga[a], Fuminori Honda[d], Ai Nakamura[d], Yoshiya Homma[d], Masashi Tokunaga[c], Dai Aoki[d], and Etsuji Yamamoto[a]

[a] *Advanced Science Research Center, Japan Atomic Energy Agency, Tokai, Ibaraki, 319-1195, Japan*
[b] *Charles University in Prague, Faculty of Mathematics and Physics, Department of Condensed Matter Physics, Ke Karlovu 5, 121 16 Prague 2, Czech Republic*
[c] *Institute for Solid State Physics, University of Tokyo, Kashiwa, Chiba 277-8581, Japan*
[d] *Institute for Materials Research, Tohoku University, Oarai, Ibaraki 311-1313, Japan*

**Abstract**
We prepared single crystals in $UCo_{1-x}Rh_xGe$ and $UIr_{1-x}Rh_xGe$ systems to establish a complex $d_{U-U}$-$T$ ($d_{U-U}$ is the shortest interatomic uranium distance and $T$ is temperature) magnetic phase diagram. This recognized a characteristic maximum in magnetic susceptibility at temperature $T_{max}$ along the $b$ axis as an important parameter. Three magnetically ordered regions can be distinguished within this scope; first a ferromagnetic region with Curie temperature < $T_{max}$, second a ferromagnetic region with Curie temperature ≈ $T_{max}$ and finally an antiferromagnetic region existing on the UIrGe side with Néel temperature < $T_{max}$.

## 1. Introduction

Magnetism of the uranium compounds is a prominent subject of condensed matter physics due to 5$f$ electrons on the boundary of localized and itinerant character. The research in this field led to the discovery of coexistence of ferromagnetism and superconductivity in the three uranium compounds $UGe_2$[1], URhGe[2] and UCoGe[3]. The close relationship between ferromagnetism and superconductivity is projected to complex $H$-$T$-$p$ phase diagrams of the given compounds. An intriguing dome of superconductivity appears along the magnetically intermediate $b$ axis in URhGe[4] in the field interval ~8-12.5 T and in the vicinity of a characteristic transition detected in the magnetization isotherm at critical magnetic field $H_R$ = 12 T. Similar broad enhancement of the superconductivity was found in the isostructural and isoelectronic UCoGe in magnetic field ~15 T along the identical axis[5]. However, high magnetic field studies detected a characteristic jump in the magnetization isotherm at a significantly higher field of almost ~46 T[6], which does not coincide with the magnetic field of the SC enhancement. Instead the $H_b$-$T$ phase diagram of UCoGe suggests a relation of the broad magnetization jump with a peculiar temperature $T_{max}$. $T_{max}$ appears as a wide maximum in the temperature dependent magnetic susceptibility along the $b$ axis at 37.5 K in UCoGe[6] in contrast to a sharp peak at 9.5 K in URhGe, which coincides with Curie temperature $T_C$[7]. $T_{max}$ was experimentally detected also in many systems like $URu_2Si_2$, $UPt_3$, $UPd_2Al_3$[8], $U_2Zn_{17}$[9], $UIr_2Zn_{20}$, and $UCo_2Zn_{20}$[10] including Ce and Yb materials[11] as a component of their magnetic phase diagrams.

A consecutive UCoGe-URhGe-UIrGe $d_{U-U}$-$T$ phase diagram, respecting Hill limit scenario[12], will be constructed based on analysis of the magnetization and heat capacity data to uncover the evolution of $T_{max}$, which remained hidden in so far performed polycrystalline studies [13-19].

## 2. Experimental part

We have prepared, for the purpose of our research, single crystals of compositions $x = 0.1$, 0.2, 0.3, 0.4 and 0.8 in UCo$_{1-x}$Rh$_x$Ge system and $x = 0.42$, 0.55, 0.57 and 0.86 in UIr$_{1-x}$Rh$_x$Ge system by Czochralski pulling in a tetra arc furnace. UCo$_{1-x}$Rh$_x$Ge single crystals were thermally treated by conventional procedure at 900°C[20, 21] and UIr$_{1-x}$Rh$_x$Ge at 1000°C[22]. An Electron-probe microanalyzer EPMA JXA-8900 (JEOL) has been used for the chemical analysis. Structural characterization was performed by single crystal X-ray diffraction using a Rigaku Rapid diffractometer, evaluated using SHELX software[23]. The temperature and magnetic field dependent magnetization were measured using a commercial magnetometer MPMS 7T (Quantum Design). Heat capacity measurements were carried out down to 1.8 K by a commercial Physical Properties Measurement System PPMS 9T (Quantum Design DynaCool).

## 3. Results

We have successfully prepared all alloy compounds in the form of high quality single crystals, which was verified by Laue method with symmetrically arranged sharp reflections in the patterns. The results of electron microprobe analysis of UIr$_{1-x}$Rh$_x$Ge are summarized in ref[22], composition of UCo$_{1-x}$Rh$_x$Ge is consistent with the stoichiometry of the melt. X-ray diffraction confirmed an orthorhombic crystal structure.

The ferromagnetic phase was detected between UCoGe and UIr$_{0.43}$Rh$_{0.57}$Ge. UIr$_{0.44}$Rh$_{0.55}$Ge – UIrGe region orders antiferromagnetically[22]. $T_C$ of the ferromagnetic phases were estimated as an inflection point of the temperature dependent magnetization and as an onset of the λ-anomaly in heat capacity data (Fig. 1). Estimated $T_C$ in UCo$_{1-x}$Rh$_x$Ge are consistent with the previous polycrystalline study. Maximal $T_C$ = 20 K detected at UCo$_{0.6}$Rh$_{0.4}$Ge significantly exceeds $T_C$ of both parent compounds[15].

We distinguished two typical examples of the λ-anomaly in the heat capacity data. A very weak λ-anomaly signaling considerably low magnetic entropy $S_{mag}$ was detected between the parent UCoGe ($S_{mag}$ = 0.003 $R$ln2)[3] and UCo$_{0.7}$Rh$_{0.3}$Ge. Only $T_C$ is gradually shifted to higher temperature with the increasing Rh content. UCo$_{0.8}$Rh$_{0.2}$Ge in Fig. 2 is typical representative of this behavior. The λ-anomaly is suddenly more pronounced at UCo$_{0.6}$Rh$_{0.4}$Ge signaling a rapid increase of the magnetic entropy. The height of the anomaly $\Delta C_p/T$ remains almost unchanged towards URhGe. UCo$_{0.2}$Rh$_{0.8}$Ge in Fig. 2 is typical representative of this behavior. The well-developed λ-anomaly also remains conserved in the ferromagnetic part of the UIr$_{1-x}$Rh$_x$Ge system[22]. Estimated $T_C$ in UCo$_{1-x}$Rh$_x$Ge will be used later for construction of the $d_{U-U}$-$T$ phase diagram. Data from the previous work will be used for construction of the UIr$_{1-x}$Rh$_x$Ge part [22].

Single crystals allowed us to carry out temperature scans of the magnetization along the $b$ axis for all compositions to study the characteristic temperature $T_{max}$. We have detected clearly $T_{max}$ between UCoGe and UCo$_{0.7}$Rh$_{0.3}$Ge. While $T_C$ increases with the increasing Rh content from UCoGe towards UCo$_{0.6}$Rh$_{0.4}$Ge, $T_{max}$ is rapidly suppressed to a lower temperature from the original 37.5 K[6] downwards 20 K in UCo$_{0.7}$Rh$_{0.3}$Ge. There is still detectable gap between $T_{max}$ and $T_C$ in UCo$_{0.7}$Rh$_{0.3}$Ge. UCo$_{0.8}$Rh$_{0.2}$Ge compound in Fig. 2 is a typical example of this behavior. Further increase of the Rh concentration to UCo$_{0.6}$Rh$_{0.4}$Ge caused the sudden development of the $T_{max}$ behavior. Originally broad maximum transforms abruptly to a sharp peak, with a position which coincides with the detected $T_C$ along the $c$ axis. This feature, typical for the parent URhGe[6], is conserved from UCo$_{0.6}$Rh$_{0.4}$Ge throughout URhGe towards the ferromagnetic boundary at UIr$_{0.43}$Rh$_{0.57}$Ge [22]. UCo$_{0.2}$Rh$_{0.8}$Ge in Fig. 2 is a typical example of this behavior.

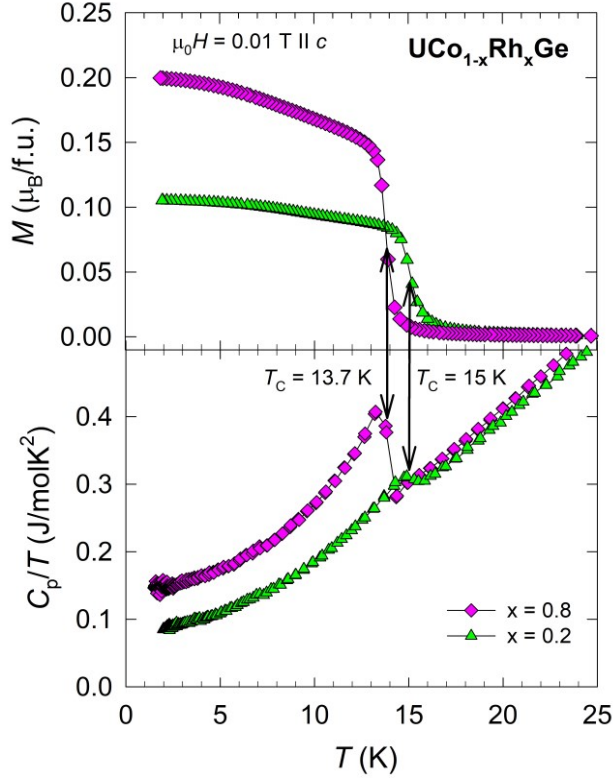

**Fig. 1.** Temperature dependent magnetization and zero magnetic field heat capacity data of $UCo_{0.8}Rh_{0.2}Ge$ and $UCo_{0.2}Rh_{0.8}Ge$. Both compounds have almost identical $T_C$ = 15 K and 13.7 K, respectively, and both are identically 20 % substituted systems on UCoGe and URhGe side of the phase diagram.

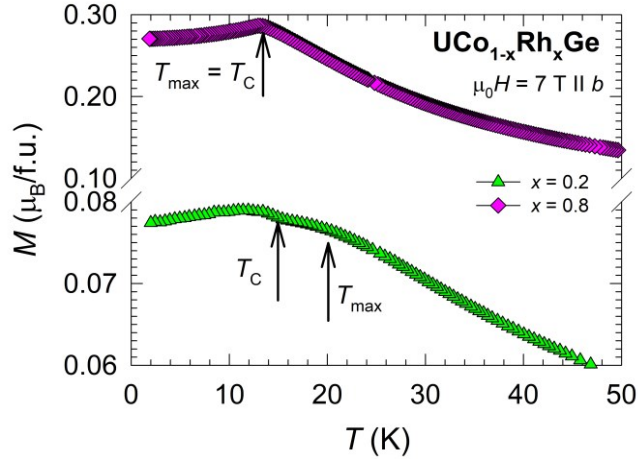

**Fig. 2.** Temperature dependence of the magnetization of $UCo_{0.8}Rh_{0.2}Ge$ and $UCo_{0.2}Rh_{0.8}Ge$. A gap of 5 K was detected between $T_C$ and $T_{max}$ in $UCo_{0.8}Rh_{0.2}Ge$. The secondary anomaly at 15 K in the magnetization data of $UCo_{0.8}Rh_{0.2}Ge$ is a projection of the easy magnetization axis because of a small imperfection of the sample alignment along the $b$ axis with respect to magnetic field. A characteristic peak-like $T_{max}$ anomaly was detected in $UCo_{0.2}Rh_{0.8}Ge$

## 4. Discussion

We have collected $T_C$, Néel temperature $T_N$ and $T_{max}$ throughout the whole system and constructed a $d_{U-U}$-$T$ phase diagram in Fig. 3. We have used $d_{U-U}$ parameter instead of $x$ because previous works have showed that lattice parameters in U$T$Ge obey Vegard's law[24] and simultaneously $d_{U-U}$ well describes evolution of the magnetism in the orthorhombic U$T$Ge compounds [18, 22, 25, 26]. Three characteristic regions can be distinguished in the $d_{U-U}$-$T$ phase diagram: The first ferromagnetic region bound by the parent UCoGe and UCo$_{0.7}$Rh$_{0.3}$Ge is characterized by $T_C < T_{max}$; Second, ferromagnetic region bound by UCo$_{0.6}$Rh$_{0.4}$Ge expands up to the boundary ferromagnetic phase in the UIr$_{1-x}$Rh$_x$Ge system with composition UIr$_{0.43}$Rh$_{0.57}$Ge. This region is specific by a rather unusual equivalence $T_C = T_{max}$. Moreover, $T_{max}$ is represented by a sharp peak-like anomaly on susceptibility data, which distinguishes it from the first ferromagnetic phase but also from the up to date so far explored uranium, and also Ce and Yb systems [8, 11]. The ferromagnetic phase is separated from the third region of the antiferromagnetic phase at composition UIr$_{0.44}$Rh$_{0.56}$Ge by a first order transition characterized by a discontinuity in all magnetic parameters, particularly by splitting of $T_{max}$ from the $T_N$. The trend $T_N < T_{max}$ is conserved towards the parent UIrGe.

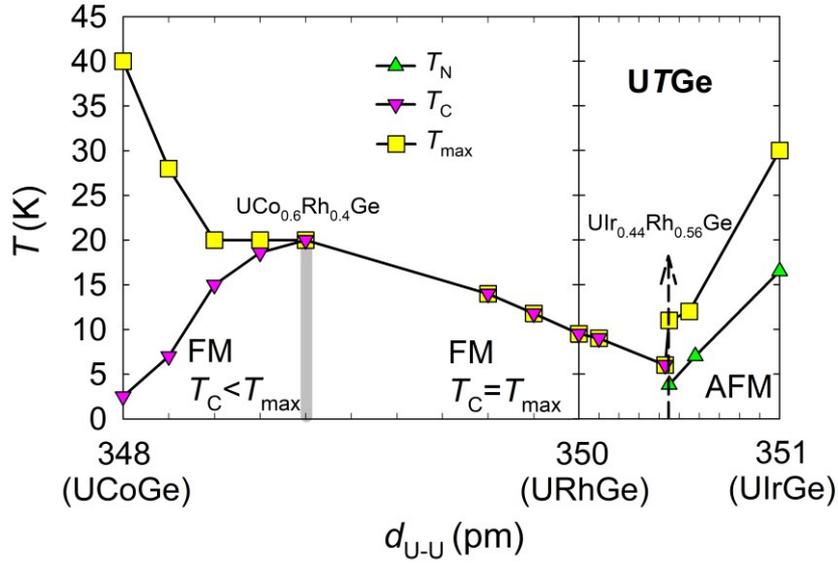

**Fig. 3.** The $d_{U-U}$-$T$ phase diagram of the isostructural and isoelectronic UCoGe-URhGe-UIrGe system. The width of the UCo$_{1-x}$Rh$_x$Ge and UIr$_{1-x}$Rh$_x$Ge panels corresponds with the nearest uranium ion distance $d_{U-U}$, assuming Vegard's law[24]. While the ordinary magnetism appears along the $c$ axis, $T_{max}$ appears exclusively along the $b$ axis. The wide gray line only tentatively marks two ferromagnetic regions with respect to $T_C$ and $T_{max}$ and does not represent a phase transition. In contrast, the dashed arrow marks the position of the ferromagnetic (FM)/antiferromagnetic (AFM) boundary separated by a first order transition[22].

The analysis of the antiferromagnetic phase has shown a relation $1.8T_N \approx T_{max}$[22]. However, any general relation between $T_C$ and $T_{max}$ was not recognized for the first ferromagnetic phase. Only rather unusual equivalence $T_C = T_{max}$ characterizes the second ferromagnetic phase between UCo$_{0.6}$Rh$_{0.4}$Ge and UIr$_{0.43}$Rh$_{0.57}$Ge. The physical phenomenon, which stands behind and is represented by $T_{max}$ is so far unclear. UCo$_{1-x}$Rh$_x$Ge and UIr$_{1-x}$Rh$_x$Ge ferromagnets were recently subject of the Rhodes-Wohlfarth analysis introducing $p_s/p_{eff}$ (spontaneous/effective moment) ratio as coming from spin fluctuation theory[27]. Both parent ferromagnetic compounds UCoGe and URhGe including alloys well fit to the scenario. It was found that $p_{eff}$ gradually decreases from 1.94 $\mu_B$/U to 1.73 $\mu_B$/U for UCoGe and UIr$_{0.43}$Rh$_{0.57}$Ge, respectively. On the other hand, $p_s$ reaches its maximum at UCo$_{0.6}$Rh$_{0.4}$Ge identical with the position of the maximum of $T_C$. Finally, UCoGe critical parameter $p_{eff}/p_s$ is ~10 times larger than that of URhGe. Thus, UCoGe is located in the limit of weak itinerant system while URhGe is an intermediate system[27]. Then, the characteristic maximum in magnetic susceptibility along the $b$ axis can be of different origin for UCoGe, URhGe and antiferromagnetic UIrGe. It can be related with instability of moments in UCoGe, while towards UIrGe it can be due to increasing antiferromagnetic correlations. The scenario of a weak itinerant character of the first FM phase is in agreement with the detected low magnetic entropy $S_{mag}$ predicted by Stoner criterion[28] in contrast with a significantly larger magnetic entropy of the compounds within the second ferromagnetic phase passing through URhGe.

## 5. Conclusions

We have successfully grown a series of single crystals throughout UCoGe-URhGe-UIrGe and constructed a unique $d_{U-U}$-$T$ phase diagram. $T_{max}$ was found as an essential characteristic temperature, which defines three regions. The ferromagnetic phases with $T_C < T_{max}$, $T_C = T_{max}$, and an AFM phase with $T_N < T_{max}$. Rhodes-Wohlfarth analysis has shown weak itinerant character of the UCoGe and related first FM phase. We have also recognized that the ferromagnetic phase with $T_C < T_{max}$ is characterized by low magnetic entropy while much larger value is suddenly detected for the FM region where $T_C = T_{max}$.

Lots of effort still needs to be invested to understand the magnetic state and sensitive magnetic interactions in the orthorhombic U$T$Ge system. Special experimental effort must be invested to determine whether a crossover can be detected between the ferromagnetic regions inside the ferromagnetic dome, suggested by gray line in Fig. 3 at $UCo_{0.6}Rh_{0.4}Ge$. The pulse field magnetization data will be published elsewhere to shed more light on the magnetism in the ferromagnetic phases of this U$T$Ge system.

**Acknowledgements**

This work was supported by JSPS KAKENHI Grant Numbers 15H05884 (J-Physics), 16K05463, JP15KK0174, JP15K05156, JP15KK0149 and JP15H05745. We would like to thank Ross H. Colman for manuscript preparation.